\documentclass[letterpaper]{emulateapj}
\submitted{Accepted by Astrophysical Journal Letters, 2008 June 17}

\newcommand{\etal}{{\it et al.}}
\newcommand{\kms}{km~s$^{-1}$}
\newcommand{\msol}{M$_{\odot}$}

\shorttitle{500 kpc HI Extension of Virgo Pair NGC4532/DDO137}

\begin{document}
\title
{A 500 kpc HI Extension of the Virgo Pair NGC4532/DDO137 Detected by
the Arecibo Legacy Fast ALFA (ALFALFA) Survey}

\author{Rebecca A. Koopmann\altaffilmark{1}$^,$\altaffilmark{2}}
\altaffiltext{1}{Union College, Department of Physics and Astronomy, Schenectady, NY 12308}
\altaffiltext{2}{National Astronomy and Ionosphere Center, 530 Space
Sciences Building, Ithaca, NY 14853. The National Astronomy and Ionosphere Center
is operated by Cornell University under a cooperative agreement with the
National Science Foundation, Cornell University.} 
\email{koopmanr@union.edu}

\author{Riccardo Giovanelli\altaffilmark{2}$^,$\altaffilmark{3}}
\altaffiltext{3}{Center for Radiophysics and Space Research and National Astronomy
and Ionosphere Center,
530 Space Sciences Building, Cornell University, Ithaca, NY 14853}
\email{riccardo@astro.cornell.edu}

\author{Martha P. Haynes\altaffilmark{2}$^,$\altaffilmark{3}}
\email{haynes@astro.cornell.edu}

\author{Brian R. Kent\altaffilmark{3}}
\email{bkent@astro.cornell.edu}

\author{Thomas J. Balonek\altaffilmark{2}$^,$\altaffilmark{4}}
\altaffiltext{4}{Department of Physics \& Astronomy, Colgate University, Hamilton, NY 13346}
\email{tbalonek@mail.colgate.edu}

\author{Noah Brosch\altaffilmark{5}}
\altaffiltext{5}{The Wise Observatory \& the School of Physics and Astronomy, Raymond \& Beverly Sackler Faculty of Exact Sciences, Tel Aviv University, Israel}
\email{noah@wise.tau.ac.il}

\author{James L. Higdon\altaffilmark{6}}
\altaffiltext{6}{Department of Physics, Georgia Southern University, Statesboro, GA 30460}
\email{jhigdon@georgiasouthern.edu}

\author{John J. Salzer\altaffilmark{7}$^,$\altaffilmark{8}}
\altaffiltext{7}{Astronomy Department, Wesleyan University, Middletown, CT 06459}
\altaffiltext{8}{Department of Astronomy, Indiana University, Bloomington, IN 47401}
\email{slaz@astro.indiana.edu}

\author{Oded Spector\altaffilmark{5}}
\email{odedspec@wise.tau.ac.il}

\begin{abstract}

We report the discovery of a $\sim$500 kpc HI extension southwest of
the Virgo Cluster HI-rich pair NGC 4532/DDO 137, detected as part of
the Arecibo Legacy Fast ALFA (ALFALFA) Survey.  The feature is the
longest and most massive HI tail structure so far found in the Virgo
Cluster and, at 1.8 Mpc from M87, the most distant from the main
concentration of the intracluster medium.  The structure is spatially
and spectrally separated into two ridges and is defined by diffuse
emission and discrete clumps of mass 2.5 - 6.8 x 10$^7$ \msol. All
emission is blue-shifted with respect to the NGC 4532/DDO 137 pair
emission.  Including diffuse emission, the structure has a total mass
of up to 7 x 10$^8$ \msol, equivalent to $\sim$10\% of the system's HI
mass. Optical $R$-band imaging finds no counterparts to a level of 26.5 mag
arcsec$^{-2}$.  The characteristics of the structure appear most
consistent with a tidal origin.

\end{abstract}


\section{Introduction}

Galaxies in clusters experience a variety of environmental
interactions that affect their evolution (see Boselli \& Gavazzi 2006 for
a review). Intracluster medium (ICM) interactions such as ram pressure
stripping (Gunn \& Gott 1972) and starvation (Larson, Tinsley, \&
Caldwell 1980) prematurely remove gaseous reservoirs. Tidal
interactions, including nearby, slower encounters in cluster and
group substructures (Toomre \& Toomre 1972), nearby high velocity encounters 
(Duc \& Bournaud 2008), and galaxy harassment (Moore \etal~
1996, 1998; Bekki \etal~2005), rearrange stellar and gaseous contents.  
These interactions potentially explain the observed gas deficiencies
(Giovanelli \& Haynes 1983, Cayatte \etal~ 1990), reduced star formation
(Kennicutt 1983; Koopmann \& Kenney 2004), and morphology-density
relation (Dressler 1980) in clusters.

Environmental interactions can produce low surface brightness
stellar and gas tails. Chung
\etal~(2007) find one-sided HI tails in seven Virgo spirals,
attributing them to the influence of the ICM.  Oosterloo \& van Gorkom
(2005) report a 110 x 25 kpc plume of HI gas extending away from the
HI-deficient Virgo spiral NGC 4388, also attributing the feature to 
ICM interaction.  Mihos \etal~(2005) find stellar streams
associated with several presumably tidal events near the Virgo Cluster core.  
Similar features have been found in Coma and Centaurus (Gregg \& West 2004).
The characteristics of tail features are related to the details of the
environmental interaction that produced them, helping to determine the 
relative importance of different interactions.  Tidal tails may
allow the formation of tidal dwarf galaxies, contributing to the 
dwarf galaxy population.

The Arecibo Legacy Fast ALFA (ALFALFA) Survey, a sensitive blind
survey of the Arecibo sky (Giovanelli \etal~2005), is providing
a complete and unbiased view of HI content and structures in
the entire Virgo cluster region. The survey has revealed the
presence of several HI clouds without optical components (Kent
\etal~2007) and a 250 kpc extended tidal 
arc emerging from the Sc galaxy NGC 4254. This structure,
encompassing the Virgo HI21 cloud (Minchin \etal~ 2007), is likely due to 
a high velocity close galaxy encounter 
(Haynes \etal~2007; Duc \& Bournaud 2008).

In this Letter, we report the detection by ALFALFA of an even larger
tidal feature associated with the Virgo Cluster pair NGC 4532/DDO 137.
This pair of SmIII/SmIV (Binggeli, Sandage, \& Tammann 1985) galaxies
is located in the Virgo B Cloud (Binggeli, Popescu, \& Tammann 1993),
6$^{\circ}$ south of the Virgo center.  NGC 4532 is the brightest Sm
cataloged in the Virgo Cluster Catalog (Binggeli, Sandage \& Tammann
1985; hereafter VCC) and has a high star formation rate (Koopmann \&
Kenney 2004) and an asymmetric stellar morphology.
The galaxies share a common HI envelope extended over 
150 kpc (Hoffman \etal~1992, 1993).  Hoffman \etal~(1999) found
that the HI envelope contains
three additional discrete HI clumps that have no optical counterparts 
as well as a significant diffuse HI component, some of
which appeared as a tail-like extension to the
southwest. We show that there is indeed an extended HI structure,
stretching $\sim$ 500 kpc beyond the pair.
 
Section~\ref{obs} describes ALFALFA and optical followup observations of the 
extended HI structure and Section~\ref{disc} 
addresses possible formation mechanisms. We assume a Virgo Cluster distance
of 16.7 Mpc (e.g., Mei \etal~ 2007) throughout.

\section{Observations and Results}
\label{obs}
\subsection{ALFALFA}
\label{hiobs}

The ALFALFA survey is mapping 7074 square degrees
of the high galactic sky visible from Arecibo, using the 7-feed
Arecibo L-band Feed Array (ALFA) on the 305 m antenna. 
The survey characteristics are documented 
by Giovanelli \etal~(2005). ALFALFA can detect HI sources with 
M$_{HI}>2\times 10^7 (W_{50}/25)^{1/2}M{_\odot}$
at the Virgo cluster distance, where $W_{50}$ is the velocity width of
the source line profile, measured at 50\% peak level, in \kms~
(Giovanelli \etal~2007; Kent \etal~2008). 
ALFALFA catalog data releases 
are accessible at http://arecibo.tc.cornell.edu/hiarchive/alfalfa/.
The structure reported in this paper was found during survey data reduction 
and further studied using a (standard)
2.4$^{\circ}$ x 2.4$^{\circ}$ data cube centered at
12$^h$34$^m$+06$^{\circ}$00$'$ [2000.0].

\begin{figure*}[h]
\centering
\includegraphics[scale=0.405]{koopmannf1a.eps} 
\includegraphics[scale=0.4]{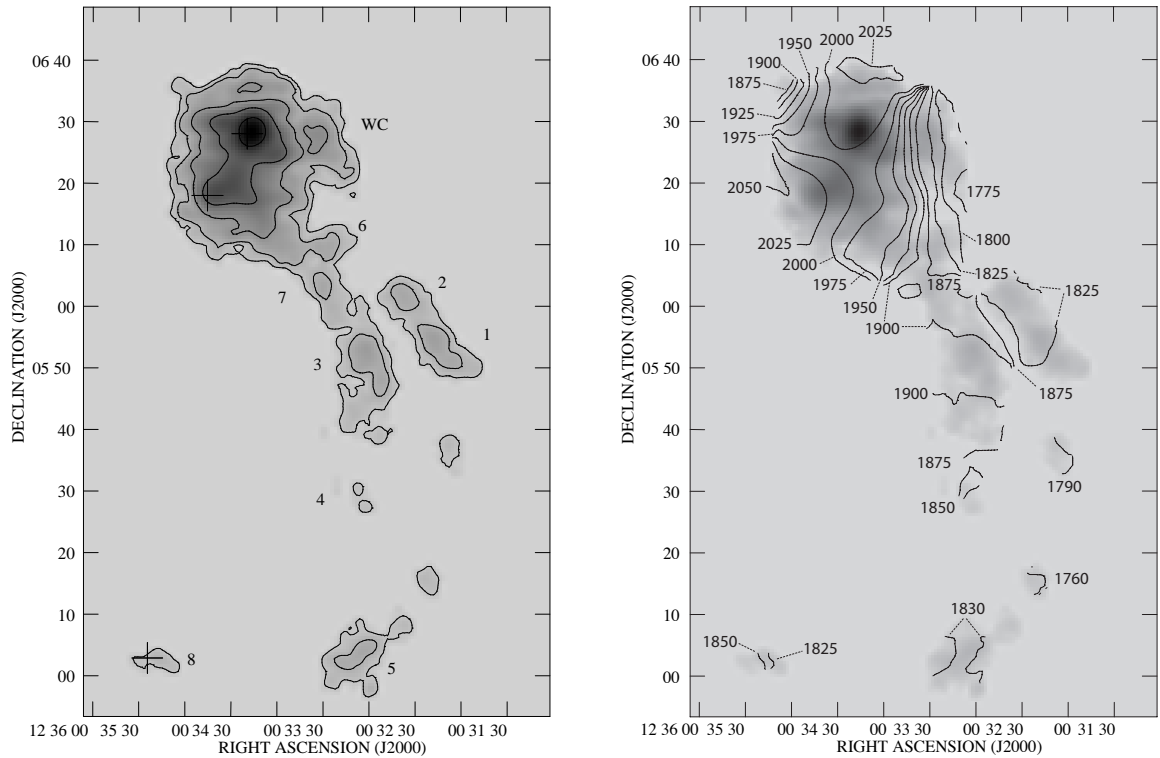}
\caption
{a. Integrated ALFALFA HI map of the NGC 4532/DDO 137 complex and extension
after convolution with a 200'' gaussian in the spatial dimension, shown
in grey-scale (using a square-root transfer function) and with
superposed contours.
Contour levels are at 0.01, 0.03, 0.12, 0.49, 2.0 M$_{\odot}$ pc$^{-2}$
(9.65 x 10$^{17}$, 3.86 x 10$^{18}$, 1.54 x 10$^{19}$, 6.18 x 10$^{19}$, 
2.47 x 10$^{20}$ cm$^{-2}$).
Clumps are numbered according to their right
ascension, as given in Table~\ref{tab1}. `WC' indicates the location of
the western clump discussed in the text. Clump \#8 coincides spatially
and spectrally with the dwarf galaxy Tololo 1232+052 and is thus a candidate
tidal dwarf galaxy.
b. Intensity weighted HI velocity field for the NGC5241/DDO137 system.
Isovelocity contours in units of km s$^{-1}$ are superposed 
on the grey-scale integrated HI distribution.
The velocity field in the HI extension is highly ordered and
all of the emission in the extension is blueshifted with respect to 
the pair.
\label{fig1}}

\end{figure*}

Figure~\ref{fig1}a shows an integrated HI map of NGC 4532/DDO 137 and 
vicinity. The extended HI envelope surrounding the pair galaxies
(Hoffman \etal~1992, 1993) is visible at upper left and the newly
detected extended HI structure emerges from the southwest.
In Figure~\ref{fig1}b, intensity weighted HI velocity contours 
are superposed on the integrated HI map. 
(For detailed HI kinematics within the envelope surrounding the pair,
see Hoffman \etal~1993, 1999).
Extensive low HI column density ($N_{\rm HI}$) gas is apparent 
throughout this region in  
several channel maps, as shown in Figure~\ref{fig2}.

We measure an HI mass of 6.0 x 10$^9$ \msol \ for the HI envelope
surrounding the pair, consistent with that of Hoffman \etal~(1999).
An additional HI mass of 1.3 x 10$^8$ \msol~ is contained within a
partially resolved clump $\sim$ 20$'$ west of NGC 4532, hereafter the
`western clump.'  The HI extension 
is defined by discrete clumps as well as
diffuse emission (Figure~\ref{fig2}).  Fluxes, velocities, and masses
of the discrete clumps are presented in Table~\ref{tab1} (see
Giovanelli \etal~2007 for details about determination of listed
parameters).  Clumps are numbered in order of their right ascension
and their positions are labeled with these numbers in
Figure~\ref{fig1}a.  Several of the clumps are marginally resolved by
the ALFALFA beam and show elongated structure.  Velocity widths of
the clumps range from 34 to 112 \kms, values similar to those of low
luminosity dwarf galaxies.  Most clumps are single-peaked.
Clump \#8 coincides spatially and in velocity with the galaxy
Tololo 1232+052, a dwarf galaxy (M$r$ =
-15) with a blue color and prominent emission lines
(SDSS; York \etal~2000), making it a tidal dwarf
candidate.

All of the emission in the HI extension is blueshifted with respect to
NGC 4532/DDO 137. Figure~\ref{fig1}b shows that the velocity field is
highly ordered.  There are two ridges separated spectrally by
about 100 \kms and spatially by about 10$'$ east-west.  The eastern
ridge is closest in velocity to the envelope surrounding NGC 4532/DDO137
and is similar in velocity
to the hook-like feature emerging northeast of the envelope (see also
Figure 2 of Hoffman \etal~1993). The western ridge, which includes the
portion that extends south and then curves toward the east, extends to
the lowest velocities in the system, overlapping spectrally with the
western clump.

The total HI mass contained within discrete clumps is 3.9 x
10$^8$ \msol.  Diffuse emission traces the ridges between the main
clumps, with an HI mass of $\sim$1x10$^8$ \msol.
An upper limit of $\sim$2 x 10$^8$ \msol~ can be placed on the
mass below the ALFALFA limiting $N_{\rm HI}$ of 3 x 10$^{18}$
cm$^{-2}$, assuming a total area of 500 x 20 kpc. Thus the total
mass in the structure has an upper limit 7 x 10$^8$ \msol, approximately 10\%
of the HI mass within the disks of the two galaxies in the pair.

\subsection{Optical Imaging}

Deep $B$- and $R$-band imaging of several fields in the HI extension
was carried out at the 40-inch telescope at Wise Observatory on the
nights of 18, 19, and 21 May, 2007, using the PI-CCD camera with a
pixel scale of 0.6 arcsec pixel$^{-1}$.  Additional fields were imaged
in $R$-band at the WIYN 0.9m at Kitt Peak Observatory on the night of
21 May, 2007, using the S2KB CCD camera with pixel scale of 0.6 arcsec
pixel$^{-1}$. Images were reduced in IRAF using standard procedures.
Wise $R$-band images from 18 May and the WIYN $R$-band images reach a
surface brightness level of 26.5 mag arcsec$^{-2}$ and 25.5 mag
arcsec$^{-2}$, respectively. Scattered light contaminates Wise
$R$-band images in several fields. The $B$-band images reach a surface
brightness level of $\sim$24.4 mag arcsec$^{-2}$, comparable to the
SDSS.  Faint galaxies visible in these fields either lack redshifts or
are at high redshift (Sloan Digital Sky Survey; York \etal~2000) and
few are good matches for the HI clump positions.  Thus, with the
exception of Tololo 1232+052, no dwarf galaxy-like sources (i.e.,
$M_{\rm B} \approx -15$, $D \approx 2-5$ kpc) are seen to coincide
with the HI clumps in Figure~\ref{fig1}a.  No obvious
extended optical emission is apparent in the smoothed images, though
scattered light limits this analysis.  
Additional optical, HI synthesis, and GALEX followup observations
are underway.

\begin{deluxetable*}{crrlrcrcl}
\tabletypesize{\scriptsize}
\tablecaption{HI Sources in the NGC4532/DDO137 Stream \label{tab1}}
\tablewidth{0pt}
\tablehead{
\colhead{Source}&
\colhead{$\alpha$}&
\colhead{$\delta$}&
\colhead{cz$_{\odot}$}&
\colhead{W$_{50}$}&
\colhead{F$_c$}&
\colhead{S/N}&
\colhead{M$_{HI}$}&
\colhead{Notes}\\
\colhead{}&
\colhead{J2000}&
\colhead{J2000}&
\colhead{(\kms)}&
\colhead{(\kms)}&
\colhead{(Jy \kms)}&
\colhead{}&
\colhead{(10$^7$ M$_{\odot}$)}&
\colhead{}
}
\startdata
Pair Complex & 12 34 20.2 & +06 27 51 & 2015 $\pm$ 1 & 163 $\pm$ \ 2 &  91.66 $\pm$ 0.08& 170 & 602 & \\ 
W Clump& 12 33 36.9& +06 26 43 & 1804 $\pm$ 9 & 103 $\pm$ 18& 1.99 $\pm$ 0.08 & 18 & 13 & \\ 
\hline\\
1&12 32 18.6 &+05 52 51 & 1833 $\pm$ 10 & 47 $\pm$ 20  & 1.03 $\pm$ 0.06 & 12.8 & 6.8 & multiple peaks \\
2&12 32 39.8 &+06 01 17 & 1803 $\pm$ 7 & 107 $\pm$ 14& 0.74 $\pm$ 0.05 &  8.4 & 4.9 & \\
3&12 33 05.3 &+05 51 00& 1915 $\pm$ 3 &  36 $\pm$ \ 8 & 0.98 $\pm$ 0.05 & 15.0 & 6.4 & SW extension\\ 
4:&12 33 05.7 &+05 26 00& 1841 $\pm$ 13 &  76 $\pm$ 25 &0.46 $\pm$ 0.06 & 5.2  & 3.0 & uncertain detection \\
5&12 33 13.4 &+05 02 47& 1826 $\pm$ 2 &  35 $\pm$ \ 5 & 0.75 $\pm$ 0.05 & 10.6 & 4.9 & multiple peaks\\
6&12 33 31.2 &+06 09 30& 1819 $\pm$ 9 & 113 $\pm$ 18 & 0.58 $\pm$ 0.05&  6.3  & 3.8 & \\ 
7& 12 33 34.0 &+06 02 59& 1872 $\pm$ 6 & 110 $\pm$ 17 & 0.98 $\pm$ 0.11&  8.6 & 6.4 & multiple peaks\\
8&12 35 20.2 &+05 02 00& 1808 $\pm$ 4 &  35 $\pm$ \ 9  & 0.38 $\pm$ 0.05&  5.4 & 2.5 & assoc. w/ Tololo1232+052?\\ 
\enddata
\end{deluxetable*}

\section{Discussion}
\label{disc}

We have discovered an extremely long (500 kpc) HI stream
of low $N_{\rm HI}$, dotted by higher density clumps appearing
as isolated HI clouds, most with no optical counterpart, apparently
associated with the galaxy pair NGC 4532/DDO 137. The
characteristics of the system are reminiscent of
those reported by Kent et al~(2007), Haynes et al~(2007), and Tripp
(2007), also found mainly in the periphery of the Virgo cluster.

The feature is to our knowledge the most extreme HI tail structure
found in a cluster, both in terms of its length and its position in
the cluster.  It is located at $\ge$1.6 times the distance from the
Virgo Cluster center as other galaxies with tail features. The
projected extent is a factor $\ge$ 14 times as large as the one-sided
HI tails discovered in the VLA Imaging of Virgo Galaxies Survey (Chung
\etal~2007). It is several times larger than the stellar tails found
by Mihos \etal~(2005) and Gregg \& West (2004) and the HI tail
described by Osterloo \& Van Gorkom (2005), and twice as large as the
feature near NGC 4254 (Haynes \etal~2007). The total HI mass is a
factor of 1.5-16 times larger than that reported for other HI tails,
although the fractional mass of $\sim$10\% of the presumed host system
is similar.  
 
The NGC 4532/DDO 137 pair is located at a projected distance of 1.8 Mpc
south of M87 and 0.5 Mpc southwest of M49 (NGC 4472).  Binggeli \etal~(1993)
identify the galaxies as members of the Virgo B cloud, which is
centered near M49 and lies at about the same distance as the more
massive Virgo A Cloud centered near M87  (Binggeli \etal~1993; Mei
\etal~2007).  The subclump contains about $\sim$1\% of the total ICM mass
in the cluster (Schindler \etal~1999) and has a spiral-rich population
with a mean line-of-sight velocity of $\sim$1040 km s$^{-1}$ and a velocity
dispersion of $\sim$500 km s$^{-1}$ (Binggeli \etal~1993). 

We estimate the ram pressure force due to the ICM at the position of
NGC 4532/DDO 137 to be 2 - 25 times smaller than the gravitational restoring
force on their ISM (following a
similar approach as Chung \etal~2007 with dynamical properties of the
galaxies given by Hoffman\etal~1999).  The extended HI envelope is
presumably less tightly bound and would be more susceptible to 
stripping. A
challenge for an ICM interpretation is the length of the feature: it
is an order of magnitude longer than other observed and simulated
(e.g., Vollmer \etal~2001; Roediger \& Br\"uggen 2008) features.  In
addition, it extends south of the pair, implying a trajectory that did
not take the pair through the densest and hottest part of the ICM, as
traced by ROSAT (B\"{o}hringer \etal ~1994) and ASCA (Shibata
\etal~2001). 

Tidal interactions naturally produce long, gas-rich tails (e.g., Toomre \&
Toomre 1972).  NGC 4532 and DDO 137 appear to be a bound pair and
could be interacting.  NGC 4532 shows other symptoms of tidal
interaction: it is optically asymmetric and has a high star formation
rate (Koopmann \etal~2004) and disturbed velocity field (Rubin \etal
1999; Chemin \etal~2005; Hoffman \etal~1999).  The HI extension
described here displays a highly ordered velocity field. However
low velocity tidal interactions between galaxies tend to produce
symmetric tails of gas and stars (e.g., Toomre \& Toomre 1972; Hibbard
\etal~2001).  
In this case no stellar tail has yet been found and the
HI extension and excess HI envelope gas not identified with the
galaxies (Hoffman \etal~1999) display strong kinematic and spatial
asymmetries.

These peculiarities could be consistent with a higher velocity
encounter with another massive galaxy.  Models of high velocity 
($\sim$ 1000 km s$^{-1}$) close galaxy
encounters (Duc \& Bournaud 2008) share some similarities to 
low velocity encounters, e.g.,
the length of the tail and the formation of dense clumps along tails,
but produce lower mass, asymmetric, gas-dominated tails.  Duc \&
Bournaud (2008) are able to reproduce the 250 kpc long HI tail
extending northward from NGC 4254 (Haynes \etal~2007) via an encounter
750 Myr ago at a speed of 1100 km s$^{-1}$ with a galaxy 50\% more
massive.
There are $\sim$10 massive (M$_B <$ -18.1) galaxies within
1.5$^{\circ}$ (440 kpc) of NGC 4532/DDO 137 and the HI extension,
including M49 (NGC 4472) and 5 other galaxies identified with Virgo B
(Binggeli \etal~1993).  NGC 4532 and DDO 137 have line-of-sight
velocities of $\sim$ 2000 km s$^{-1}$ (2$\sigma \sim$ 1000 km s$^{-1}$
greater than Virgo B mean) so that a high speed encounter with a B
member is possible.  As argued by Duc \& Bournaud (2008), the
perturber may be further away; a galaxy moving at 1000 km s$^{-1}$ can
travel a projected distance of $\sim$ 1 Mpc in 1 Gyr.  We note that
ALFALFA observations, to date complete to +4$^{\circ}$00$'$, show no
other extended HI features associated with other galaxies in the
vicinity.

Based upon the available models, we suggest that the structure
associated with NGC 4532/DDO 137 is most consistent with a tidal
interaction, possibly a high velocity encounter.  Determining the
exact nature of these very long HI tails and the extended HI envelope
will require detailed simulation of the system in the entire Virgo
Cluster environment, an exercise outside the scope of this paper.

\acknowledgments
The authors acknowledge helpful comments from an anonymous referee.
RG, MPH, and BRK acknowledge support from grants AST--0307661,
AST--0435697 and AST--0607007 and by the Brinson Foundation.  RAK and
TJB thank NAIC for partial support. RAK, TJB, and NB are grateful to the
NAIC and the Cornell University Astronomy Department for its hospitality 
to them as sabbatic visitors and to their home institutions for sabbatic
support.



\begin{figure}
\centering
\caption
{Smoothed (6$\arcmin$ FWHM) ALFALA channel maps 
of the NGC 4532/DDO 137 system showing extensive low $N_{\rm HI}$ gas.  
A square-root transfer function has been used, and the color-bar shows 
flux density in units of mJy beam$^{-1}$. The positions of NGC 4532 and 
DDO 137 are indicated by crosses. The scale bar corresponds to 
100 kpc at the assumed distance of 16.7 Mpc.
\label{fig2}}

\end{figure}


\clearpage

\begin{references}
\reference{} Bekki, K., Koribalski, B. S., \& Kilborn, V. A. 2005, MNRAS 363, L21
\reference{} Binggeli, B., Sandage, A., \& Tammann, G. A. 1985, AJ, 90, 1681
\reference{} Binggeli, B., Popescu, C. C., \& Tammann, G. A , 1993 A\&A Supp, 98, 275
\reference{} B\"{o}hringer, H., Briel, U. G., Schwarz, R. A., Voges, W., Hartner, G., \& Trumper, J. 1994, Nature, 368, 828
\reference{} Boselli, A.  \& Gavazzi, G. 2006, PASP, 118, 517
\reference{} Cayatte, V., van Gorkom, J. H., Balkowski, C., \& Kotanyi, C. 1990, AJ, 100, 604
\reference{} Chemin, L. \etal~2005, A\&A, 436, 469
\reference{} Chung, A., van Gorkom, J. H., Kenney, J. D. P., \& Vollmer, B. 2007, ApJ, 659, 115
\reference{} Dressler, A. 1980, ApJ, 236, 351
\reference{} Duc, P.-A. \& Bournaud, F. 2008, ApJ, 673, 787
\reference{} Giovanelli, R. \& Haynes, M. P. 1983, AJ 88, 881
\reference{} Giovanelli, R. et al. 2007, AJ, 133, 2569
\reference{} Giovanelli, R. et al. 2005, AJ, 130, 2613
\reference{} Gregg, M.D. \& West, M. J. 2004, in IAU Symp. 217, Recycling Intergalactic and 
Interstellar Matter, ed. P.-A. Duc, J. Braine, \& E. Brinks (San Francisco: ASP), 70
\reference{} Gunn, J. E. \& Gott, J. R. I. 1972, ApJ, 176, 1
\reference{} Haynes, M. P., Giovanelli, R., \& Kent, B. R. 2007, ApJL, 665, L19
\reference{} Hibbard, J., van der Hulst, J., Barnes, J., \& Rich, R. 2001, 
AJ, 122, 2969
\reference{} Hoffman, G. L., Salpeter, E. E., Lamphier, C., \& Roos, T. 1992, ApJ, 388, L5
\reference{} Hoffman, G. L., Lu, N. Y., Salpeter, E. E., Farhat, B., Lamphier C., \& Roos, T. 1993, AJ, 106, 39
\reference{} Hoffman, G. L., Lu, N. Y., Salpeter, E. E., \& Connell, B. M. 1999, AJ, 117, 811
\reference{} Kennicutt, R. C. 1983, AJ, 88, 483
\reference{} Kent, B. R. \etal~ 2007, ApJL, 665, L15
\reference{} Kent, B. R. \etal~ 2008, ApJ, submitted
\reference{} Koopmann, R. A. \& Kenney, J. D. P. 2004, 613, 851
\reference{} Larson, R. B., Tinsley, B.M. \& Caldwell, C. N. 1980, ApJ, 237, 692
\reference{} Mei, S. \etal~ 2007, ApJ, 655, 144
\reference{} Mihos, J. C., Harding, P., Feldmeier, J., \& Morrison, H. 2005, ApJ, 631, L41
\reference{} Minchin, R. \etal~ 2007, ApJ, in press
\reference{} Moore, B., Katz, N., Lake, G., Dressler, A., \& Oemler, A. 1996, Nature, 379, 613
\reference{} Moore, B., Lake, G., Katz, N. 1998, ApJ, 495, 139
\reference{} Oosterloo, T. \& van Gorkom, J. 2005, A\&A, 437, L19
\reference{} Roediger, E. \& Br\"uggen 2008, MNRAS, in press
\reference{} Schindler, S., Binggeli, B., \& B\"{o}hringer, H 1999, A\&A, 343, 420
\reference{} Shibata, R. \etal~2001, ApJ, 549, 228
\reference{} Toomre, A. \& Toomre, J. 1972, ApJ, 178, 623
\reference{} Tripp, T. M. 2007, in IAU Symp. 244, Dark Galaxies and Lost Baryons, ed. J. I. Davies
\& M. D. Disney (Cambrige: Cambridge University Press), in press
\reference{} Vollmer, B., Cayatte, V., Balkowski, C., \& Duschl, W. J. 2001, ApJ, 561, 708
\reference{} York, D. G. \etal~2000, AJ, 120, 1579
\end{references}
\end{document}